\documentclass[%
 preprint,
 showpacs,
 showkeys,
 amsmath,
 amssymb,
 aps,
 pra,
  longbibliography,
 ]{revtex4-1}
 \usepackage{graphicx}

\begin{document}

\sloppy

\title{Experimental Evidence of Quantum Randomness Incomputability}

\author{Cristian S. Calude}
\email{cristian@cs.auckland.ac.nz}
\homepage{http://www.cs.auckland.ac.nz/~cristian}

\author{Michael J. Dinneen}
\email{mjd@cs.auckland.ac.nz}
\homepage{http://www.cs.auckland.ac.nz/~mjd}

\affiliation{Department of Computer Science, University of Auckland,
Private Bag 92019, Auckland, New Zealand}

\author{Monica Dumitrescu}
\affiliation{Faculty of Mathematics and Computer Science, University of Bucharest,
Str. Academiei 14, 010014 Bucharest, Romania}
\email{mdumi@fmi.unibuc.ro}
\homepage{http://fmi.unibuc.ro/ro/dumitrescu\_monica}

\author{Karl Svozil}
\affiliation{Institute for Theoretical Physics, {\it Vienna} University of Technology,
Wiedner Hauptstrasse 8-10/136, 1040 {\it Vienna},  Austria}
\email{svozil@tuwien.ac.at}
\homepage{http://tph.tuwien.ac.at/~svozil}

\date{\today}

\begin{abstract}
In contrast with software-generated randomness (called pseudo-randomness), quantum randomness is provable incomputable, i.e.\ it is not exactly reproducible by any algorithm.
  We provide experimental evidence of  incomputability --- an asymptotic property --- of quantum randomness by
  performing finite tests of randomness inspired by algorithmic information theory.
\end{abstract}

\preprint{CDMTCS preprint nr. 382/2010}
\maketitle

\section{Quantum indeterminacy}

The irreducible indeterminacy of individual quantum processes postulated by
Born~\cite{born-26-1,born-26-2,zeil-05_nature_ofQuantum}
implies that there exist physical ``oracles,'' which are capable to effectively produce outputs which are incomputable.
Indeed, quantum indeterminism has been proved~\cite{2008-cal-svo} under some  ``reasonable'' side assumptions implied by
Bell-, Kochen-Specker- and Greenberger-Horne-Zeilinger-type theorems.
Yet, as quantum indeterminism is nowhere formally specified,
it is important to investigate which (classes of) measurements lead to randomness,
what are the reasons for possible distinctions,
whether or not the kinds of randomness ``emerging'' in different classes of quantum measurements are ``the same'' or ``different,''
and what are the phenomenologies or signatures of these randomness classes.
Questions about ``degrees of (algorithmic) randomness" are studied in algorithmic information theory.
Here are just four types, among an infinity of others:
(i) standard pseudo-randomness produced by
software  like {\it Mathematica} or {\it Maple}  which
are not only Turing computable but cyclic;
(ii) pseudo-randomness
produced by software which is Turing computable but not cyclic
(e.g., digits of $\pi$, the ratio between the circumference and the diameter of an ideal circle, or Champernowne's constant);
(iii) Turing incomputable, but not
algorithmically random;
(iv) algorithmically random~\cite{MartinLöf1966602,chaitin:01,calude:02}.
One can ask: in which
of these four classes do we find quantum randomness?
Operationally, in the extreme form, Born's postulate could be interpreted to allow for the production of  ``random'' finite strings;
hence quantum randomness could be of type (iv).
(Here the quotation mark refers to the fact that randomness for finite strings is too ``subjective''
to be meaningful for our analysis.
The  legitimacy of the experimental approach comes from  characterizations of  random sequences
in terms of the degrees of incompressibility of their finite prefixes.~\cite{MartinLöf1966602,chaitin:01,calude:02}.)
A sequence which is not algorithmically random but Turing incomputable can, for instance, be obtained from an algorithmically random sequence $x_{1}x_{2}\cdots x_{n}\cdots $ by inserting a 0 in between any adjacent original bits, i.e.\ obtaining the sequence $x_{1}0x_{2}0\cdots 0x_{n}0\cdots $ This transformation destroys algorithmic randomness because obvious correlations have appeared; Turing incomputability is invariant under this transformation because a copy of the original sequence is embedded in the new one.
Yet much more subtler correlations among subsequences of Turing incomputable
sequences may exist, thus making them compressible and algorithmically nonrandom.
There is no {\em a priori} reason to interpret Born's indeterminism by its strongest formal expression; i.e.,
in terms of algorithmic randomness.

Quantum randomness produced by quantum
systems which have no classical interpretation is provable~\cite{2008-cal-svo} Turing
incomputable. More precisely, if the experiment would run under ideal
conditions ``to infinity,'' the resulting infinite sequence of bits would be
Turing incomputable; i.e., no Turing machine (or algorithm) could
reproduce exactly this infinite sequence of digits. This result has many consequence; here is one example.
The experiment could produce a billion of 0s, but not all bits produced will be 0.
A stronger form of incomputability holds true:  every Turing machine (or algorithm) can reproduce exactly
only finitely many scattered digits of that infinite sequence.
Yet this proof stops short of showing that the sequence produced by such a quantum
experiment is algorithmically random; i.e., it is unknown whether or not such
a sequence is or is not algorithmically random.
One of the strategies toward answering this question is to empirically perform tests
``against'' the algorithmic randomness hypothesis.

Our (more modest) aim is to present tests capable of  distinguishing computable
from incomputable sources of ``randomness'' by examining (long, but) finite prefixes of infinite sequences.
Such differences are guaranteed to exist by ~\cite{2008-cal-svo}, but,
because computability is an asymptotic property,
there was no guarantee that finite tests can ``pick'' differences in the prefixes we have analyzed.

\section{Tests of experimental quantum indeterminacy}

Based on Born's postulate, several quantum random number generators based on beam splitters
have recently been proposed and
realized~\cite{svozil-qct,rarity-94,zeilinger:qct,stefanov-2000,0256-307X-21-10-027,wang:056107,fiorentino:032334,svozil-2009-howto}.
In what follows a detailed analysis of bit strings of length $2^{32}$ obtained by
two such quantum random number generators will be presented --- the first analysis of
a set of quantum bits of this size (the size correlates well with  the square root of the cycle length used by cyclic pseudo-random generators; randomness properties of longer strings generated in this way  are impaired).
We will  compare  the performance of quantum random number generators
with  software-generated number generators on randomness inspired by algorithmic information theory
(which complement  some commonly used statistical tests implemented in ``batteries'' of test suites such as, for instance,
{\em diehard}~\cite{diehard},
{\it NIST}~\cite{Rukhin-nist},
or {\it TestU01}~\cite{1268777}).
The standard test suites are often based on
tests which are not designed for physical random number generators,
but rather to quantify the quality of the cyclic pseudo-random numbers generated by algorithms.
As we would like to separate ``truly'' random sequences from software-generated random sequences,
the emphasis is on the former type of  tests.

The tests based on algorithmic information theory directly
analyze randomness, and thus the strongest possible form of incomputability. They differ from tests employed in the standard  randomness batteries as they depend on irreducible algorithmic information content,
which is  constant for algorithmic pseudo-random sequences.
Some tests are related to each other, as for instance  sequences which are not Borel normal (cf. below) could be algorithmically compressed;
the analysis of results helps understanding subtle differences at the edge of incomputability/algorithmic randomness. All tests depend on the size of the analyzed strings; the legitimacy of our approach is given by the fact that algorithmic randomness of an infinite sequence can be ``uniformly read'' in its prefixes (cf. \cite{calude:02}).

\section{Data sources}

The analyzed quantum data consist of
10 quantum random strings generated with the commercially available {\it Quantis} device~\cite{Quantis},
based on research of a group in Geneva~\cite{stefanov-2000},
as well as
10 quantum random strings generated by the {\it Vienna} {\it IQOQI} group~\cite{Vienna}.
The pseudo-random data consist of
10 pseudo-random strings  produced by {\it Mathematica}~6~\cite{MRG}, and
10 pseudo-random strings  produced by  {\it Maple}~11~\cite{MAPLE},
as well as
10 strings of $2^{32}$ bits from the binary expansion of $\pi$
obtained from the {U}niversity of {T}okyo's supercomputing center~\cite{pi}.

The signals of the {\it Quantis} device are generated
by a light emitting diode producing photons which are then transmitted toward a beam splitter (a semi-transparent
mirror) and two successive  single-photon detectors (detectors with single-photon resolution)
to record the outcomes associated with the symbols ``$0$'' and ``$1$,'' respectively~\cite{Quantis}.
Due to hardware imbalances which are difficult to overcome at this level,
Quantis processes this raw data by un-biasing the sequence by a von Neumann type normalization:
The biased raw sequence of zeroes and ones is partitioned into
fixed subsequences of length two; then the even parity sequences ``$00$'' and ``$11$'' are discarded,
and only the odd parity ones ``$01$'' and ``$10$'' are kept.
In a second step, the remaining sequences are mapped into the single symbols $01 \mapsto 0$ and  $10 \mapsto 1$,
thereby extracting a new unbiased sequence at the cost of a loss of original bits~\cite[p. 768]{von-neumann1}.

This normalization method requires that the events are (temporally) uncorrelated and thus independent.
(For the sake of a simple counterexample, the von Neumann normalization of
the sequences $010101 \cdots $ or $1100110011 \cdots $ are the constant-0 sequence $000\cdots$ and the empty sequence.)
Under the independence hypothesis,  the normalized sequences  are Borel normal~\cite{borel:09}; e.g., all finite subsequences
of length $n$ occur with their expected asymptotic frequencies $2^{-n}$.
(Alas, see \cite{Hertling:jucs_8_2:simply_normal_numbers_to}
for some pitfalls when transforming such sequences.)

The signals of the Vienna Institute for Quantum Optics and Quantum
Information (IQOQI)
group were generated with photons from a weak blue LED light source which impinged on a
beam splitter
without any polarization sensitivity with two output ports associated
with the codes ``0'' and ``1,'' respectively~\cite{zeilinger:qct}.
There was {\em no} pre- or post-processing of the raw data stream,
in particular no von Neumann normalization as discussed for the Quantis device;
however the output was constantly monitored
(the exact method is subject to a patent pending).
In very general terms, the setup needs to be running for at least one
day to reach a stable operation.
There is a regulation mechanism which keeps track of the bias between
``0'' and ``1,''
and  tunes the random generator for perfect symmetry.
Each data file was created in one continuous run of the device
lasting over hours.

We have employed the {\it extended cellular automaton generator} default of {\it Mathematica}~6's pseudo-random function.
It is based on a particular five-neighbor rule, so each new cell depends on five nonadjacent cells from the previous step~\cite{MRG}.
{\it Maple}~11 uses a Mersenne Twister algorithm to generate a random pseudo-random output~\cite{MAPLE}.

\section{Testing incomputability and randomness}

The  tests  we performed can be grouped into: (i) two tests based on algorithmic information theory,
(ii) statistical tests involving frequency counts (Borel normality test), (iii)
a test based on Shannon's information theory, and (iv) a test based on random walks.

In Figures~\ref{fig:example3}--\ref{fig:example5} the graphical representation of the results is rendered in terms of
box-and-whisker plots, which characterize
groups of numerical data through five characteristic summaries:
test minimum value, first quantile (representing one fourth of the test data),
median or second quantile (representing half of the test data),
third quantile (representing three fourths of the test data),
and test maximum value. Mean and standard deviation of the data representing the results of the tests are calculated.
Tables containing the experimental data and the programs used
to generate the data can be downloaded from our extended paper~\cite{CDMTCS372}.


\subsection{Book stack randomness test}

The {\em book stack}
(also known as ``move to front'')
test~\cite{MR2099021,MR2162569} is based on the fact that compressibility is a symptom of less randomness.

\begin{figure}[htbp] 
   \centering
   \includegraphics[width=5in]{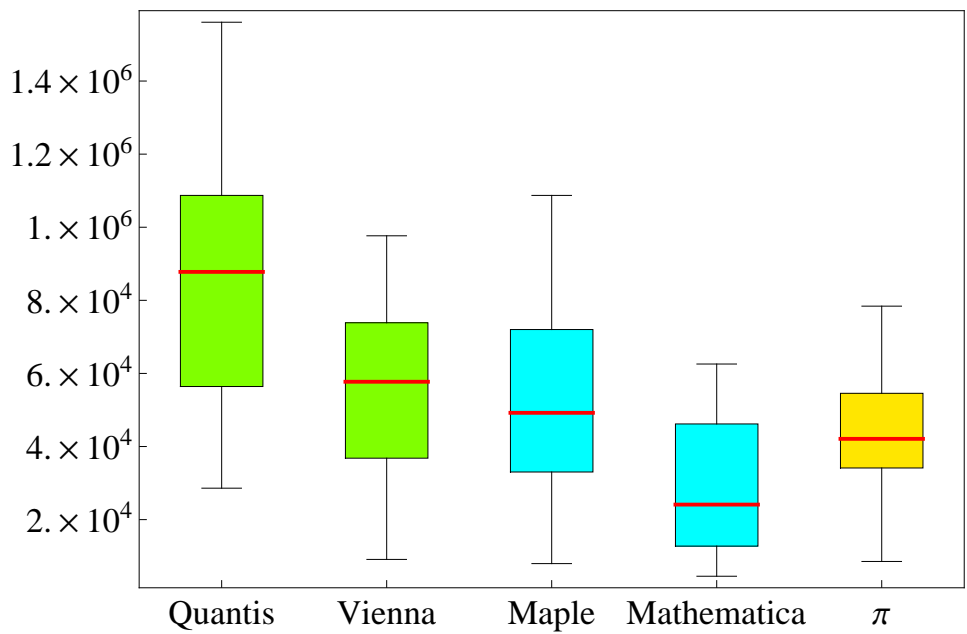}
   \caption{(Color online) Box-and-whisker plot for the results of  the ``book stack'' randomness test.}
   \label{fig:example3}
\end{figure}

The results, presented in
Figure~\ref{fig:example3} and Table~\ref{tab:3}, are derived from the original count, the count after the application
of the transformation, and the difference.  The key metric for this test is
the count of ones after the transformation.  The book stack encoder does
not compress data but instead rewrites each byte with its index (from
the top/front) with respect to its input characters being
stacked/moved-to-front.
Thus, if a lot of repetitions occur (i.e., a symptom of
non-randomness), then the output  contains more zeros than ones due to
the sequence of indices generally being smaller numerically.

\begin{table}
\caption{Statistics for the results of  the ``book stack'' randomness
test.}\label{tab:3}
\begin{center}
\begin{tabular}
[c]{ c c c c c c c c }%
\hline\hline
Descriptive statistics & min & Q1 & median & Q3 & max & mean & sd\\\hline
Maple & 7964 & 34490 & 49220 & 69630 & 108700 & 53410 & 33068.58\\
Mathematica & 4508 & 13020 & 24110 & 43450 & 62570 & 27940 & 19406.03\\
Quantis & 28600 & 60480 & 87780 & 106700 & 156100 & 89990 & 41545.76\\
Vienna & 9110 & 38420 & 57720 & 73220 & 97660 & 53860 & 27938.92\\
$\pi$ & 8551 & 35480 & 42100 & 52870 & 78410 & 41280 & 20758.46\\\hline\hline
\end{tabular}
\end{center}
\end{table}

\subsection{Solovay-Strassen probabilistic primality test}

The second algorithmic test, based on the {\em Solovay-Strassen probabilistic primality test},
uses  Carmichael (composite) numbers which are ``difficult'' to  factor, to determine the quality of randomness
by computing how fast the probabilistic primality test reaches the verdict ``composite'' \cite{solovay:84,ch-schw-78}.
All Carmichael numbers less than $10^{16}$ have been used~\cite{Pinch,Pinch07}.

To test whether  a positive integer
$n$ is prime, we take $k$ natural numbers uniformly distributed between 1
and $n - 1$, inclusive, and, for each one $i$, check whether the predicate
$W (i, n)$ holds.  If this is the case we say that
``$i$ is a witness of $n$'s compositeness''.
If $W (i, n)$ holds for at least one $i$ then $n$ is
composite; otherwise, the test is inconclusive, but in this case if one declares $n$ to be  prime then the
 probability to be wrong is smaller than $ 2^{-k}$.

 This is due to the fact that
 at least half  $i$'s from $1$ to $ n - 1$ satisfy $W (i, n)$  if $n$ is indeed composite,
 and {\it none} of them satisfy $W (i, n)$ if $n$ is prime~\cite{solovay:84}.
Selecting $k$ natural numbers  between 1
and $n - 1$ is the same as choosing a binary string $s$ of length $n-1$ with $k$ $1$'s
such that the $i$th bit is 1 iff $i$ is selected.
Ref.~\cite{ch-schw-78} contains a  proof that, if $s$ is a long enough   algorithmically random binary string,
then $n$ is prime iff $Z(s,n)$ is true,
where $Z$ is a predicate constructed directly from conjunctions of negations of  $W$~\footnote{
In fact, every  ``decent'' Monte Carlo simulation algorithm in  which  tests are chosen according to an
algorithmic random string produces a result
which is not only true with high probability, but {\it rigorously correct}~\cite{MR757602}.}.

A Carmichael number is a composite positive integer $k$ satisfying the congruence $b^{k-1} \equiv 1 ({\rm mod } \, k)$
for all integers $b$ relative prime to $k$.
Carmichael numbers are composite, but are difficult to factorize and thus are ``very similar'' to primes;
they are sometimes called pseudo-primes.
Carmichael numbers can fool Fermat's primality test, but less   the Solovay-Strassen test.
With increasing values, Carmichael numbers
become ``rare''~\footnote{There are 1,401,644 Carmichael numbers in the interval $[1, 10^{18}]$.}.

\begin{figure}[htbp] 
   \centering
   \includegraphics[width=5in]{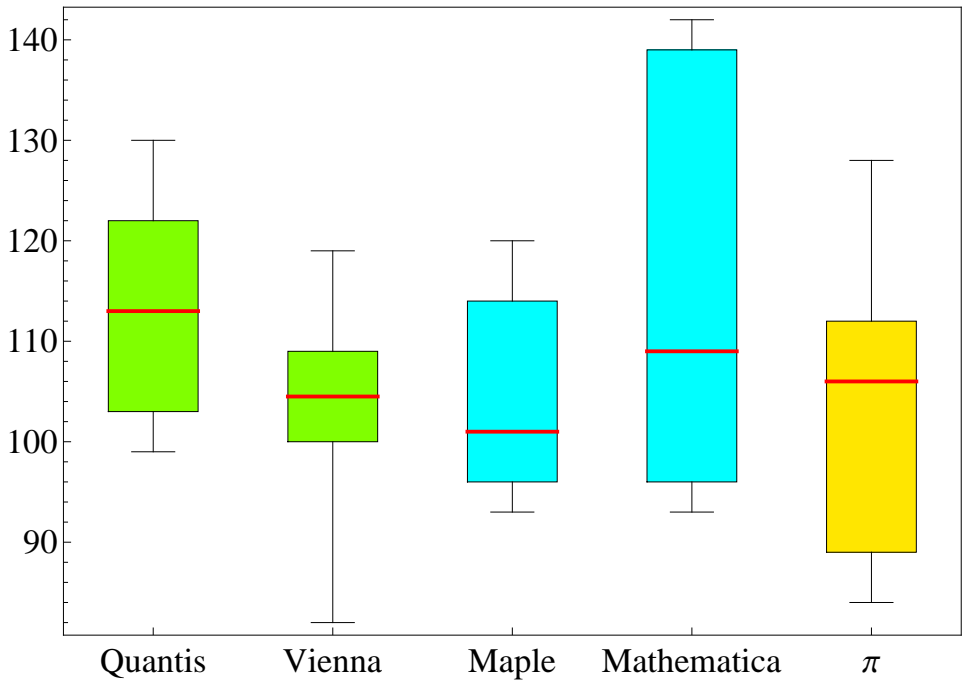}
   \caption{(Color online) Box-and-whisker plot for the results based on the Solovay-Strassen probabilistic primality test.}
   \label{fig:example4}
\end{figure}

The fourth test uses Solovay-Strassen probabilistic primality test for
Carmichael numbers (composite) with prefixes of the sample strings as the binary
string $s$. We used the Solovay-Strassen  test for all Carmichael numbers less
than $10^{16}$---computed in Ref.~\cite{Pinch,Pinch07}---with numbers selected according to increasing prefixes of each sample string till the algorithm returns
a non-primality verdict. The metric is given by the length of the sample used to
reach the correct
verdict of non-primality for all of the 246683 Carmichael numbers less than
$10^{16}$.  [We started with $k=1$ tests (per each Carmichael number) and increase $k$
until the metric goal is met; as $k$ increases we always use new bits (never
recycle) from the sample source strings.]
The results are presented in
Figure~\ref{fig:example4} and Table~\ref{tab:4}.

\begin{table}
   \caption{Statistics for the results based on the
Solovay-Strassen probabilistic primality test.}\label{tab:4}
\begin{center}
\begin{tabular}
[c]{ c c c c c c c c }%
\hline\hline
Descriptive statistics & min & Q1 & median & Q3 & max & mean & sd\\\hline
Maple & 93.0 & 96.0 & 101.0 & 113.5 & 120.0 & 104.9 & 10.57723\\
Mathematica & 93.0 & 97.0 & 109.0 & 132.3 & 142.0 & 113.5 & 19.60867\\
Quantis & 99.0 & 103.3 & 113.0 & 121.3 & 130.0 & 112.6 & 10.66875\\
Vienna & 82.0 & 100.3 & 104.5 & 109.0 & 119.0 & 103.5 & 11.03781\\
$\pi$ & 84.0 & 91.75 & 106.0 & 110.8 & 128.0 & 104.7 & 10.66875\\\hline  \hline
\end{tabular}
\end{center}
\end{table}

\subsection{Borel normality test}

{\em Borel normality} --- requesting that
every  binary string appears in the sequence with the correct probability $2^{-n}$ for a string of length $n$ ---
served as the first mathematical definition of randomness~\cite{borel:09}.
A sequence is (Borel) normal if every  binary string appears in the sequence with the right probability
(which is $2^{-n}$ for a string of length $n$).
A  sequence is normal if and only  it  is incompressible by any information lossless finite-state compressor~\cite{ZL:1978}, so  normal sequences are  those sequences  that appear random to any finite-state machine.

Every algorithmic random infinite sequence is Borel normal~\cite{DBLP:conf/dlt/Calude93}.
 The converse implication is not true:
there exist computable normal sequences (e.g.,  Champernowne's constant).

Normality is invariant under finite variations: adding, removing, or changing a finite number of bits in any normal sequence leaves it normal. Further, if a sequence satisfies the normality condition
for strings of length $n+1$, then it also satisfies normality for strings of length $n$, but the converse is not true.

Normality was transposed to  strings in Ref.~\cite{DBLP:conf/dlt/Calude93}. In this process one  has to replace limits with inequalities. As a consequence, the above two properties, which are valid for sequences, are no longer true for strings.

For any fixed  integer $m > 1$, consider the alphabet $B_{m} = \{0,1\}^{m}$ consisting of all binary strings of length $m$,
 and for
every $1 \leq i \leq 2^{m}$ denote by $N_{i}^{m}$ the
 number of occurrences of the lexicographical $i$th binary string of length $m$ in the string $x$ (considered over the alphabet $B_{m}$).
 By $|x|_{m}$ we denote the length of $x$.
 A string $x$ is Borel normal if
  for every natural $1 \leq m \leq \log_{2}\log_{2} |x|,$
  \[
\left| \frac{N_{j}^{m}(x)}{|x|_{m}} -  2^{-m} \right| \leq
\sqrt\frac{\log_{2}|x|}{|x|}\raisebox{0.5ex}{,}\]
for every $1 \leq j \leq 2^{m}$.
In Ref.~\cite{DBLP:conf/dlt/Calude93}
it is shown that almost all algorithmic random strings are Borel normal.

\begin{figure}[htbp] 
   \centering
   \includegraphics[width=5in]{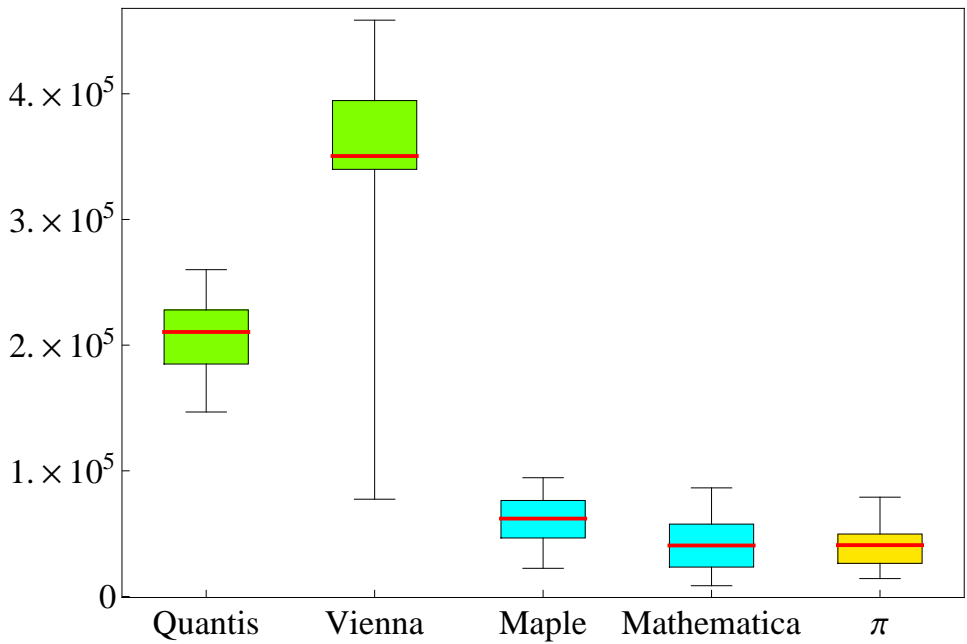}
   \caption{(Color online) Box-and-whisker plot for the results for tests of the Borel normality property.}
   \label{fig:example1}
\end{figure}

In the first test
we  count the maximum, minimum and difference of non-overlapping occurrences of  $m$-bit ($m=1,\ldots , 5$) strings  in
each sample string.   Then we tested the Borel normality property for each sample string and found that
almost all strings pass the  test,  with some notable exceptions.
We found that several of the Vienna sequences failed the expected count range for
$m=2$ and a few of the Vienna sequences were outside the expected range for $m=3$
and $m=4$ (some less then the expected minimum count and some more than the expected
maximum count).  The only other bit sequence that was outside the expected range
count was one of the Mathematica sequences that had a too big of a count for $k=1$.
Figure~\ref{fig:example1} depicts a box-and-whisker plot of the
results. This is followed by statistical (numerical) details
in Table~\ref{tab:1}.

\begin{table}
\caption{Statistics for the results for tests of the Borel normality property.}\label{tab:1}
\begin{center}
\begin{tabular}
[c]{c c c c c c c c c c c c c c c }%
\hline\hline
Descriptive statistics & min & Q1 & median & Q3 & max & mean & sd\\\hline
Maple & 22430 & 47170 & 61990 & 76130 & 94510 & 60210 & 21933.52\\
Mathematica & 8572 & 25500 & 40590 & 55650 & 86430 & 41870 & 23229.77\\
Quantis & 146800 & 185100 & 210500 & 226600 & 260000 & 207200 & 33515.65\\
Vienna & 77410 & 340200 & 350500 & 392500 & 260000 & 337100 & 103354.3\\
$\pi$ & 14260 & 28860 & 40880 & 47860 & 79030 & 40220 & 17906.21\\\hline\hline
\end{tabular}
\end{center}
\end{table}

\subsection{Test based on Shannon's information theory}

\begin{figure}[htbp] 
   \centering
   \includegraphics[width=5in]{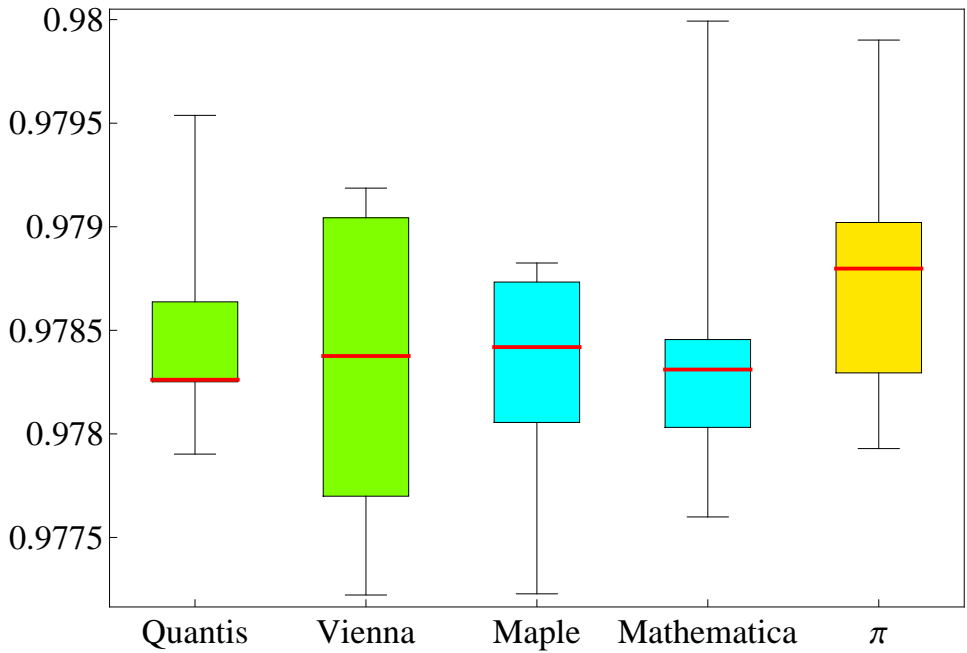}
   \caption{(Color online) Box-and-whisker plot for average results in ``sliding window'' estimations of the Shannon entropy.}
   \label{fig:example2}
\end{figure}

The next test
computes  ``sliding window'' estimations of the Shannon entropy $L_n^1, \ldots ,L_n^t$ according to the method described in
\cite{Wyner}: a smaller  entropy is a symptom of  less randomness.
The results are presented in
Figure~\ref{fig:example2} and Table~\ref{tab:2}.

\begin{table}
\caption{Statistics for average results in ``sliding window''
estimations of the Shannon entropy.}\label{tab:2}
\begin{center}
\begin{tabular}
[c]{c c c c c c c c c c c c c c c }%
\hline\hline
Descriptive statistics & min & Q1 & median & Q3 & max & mean & sd\\\hline
Maple & 0.9772 & 0.9781 & 0.9784 & 0.9787 & 0.9788 & 0.9783 & 0.0005231617\\
Mathematica & 0.9776 & 0.9781 & 0.9783 & 0.9785 & 0.9800 & 0.9783 & 0.0006654936\\
Quantis & 0.9779 & 0.9783 & 0.9783 & 0.9786 & 0.9795 & 0.9784 & 0.0004522699\\
Vienna & 0.9772 & 0.9777 & 0.9784 & 0.9790 & 0.9792 & 0.9783 & 0.0006955834\\
$\pi$ & 0.9779 & 0.9784 & 0.9788 & 0.9790 & 0.9799 & 0.9788 & 0.0006062724\\\hline\hline
\end{tabular}
\end{center}
\end{table}

\subsection{Test based on random walks}

A symptom of non-randomness of a string is detected when the plot generated by viewing a sample sequence as
a 1D random walk meanders ``less  away''
from the starting point (both ways); hence the max-min range is the metric.

\begin{figure}[htbp] 
   \centering
   \includegraphics[width=5in]{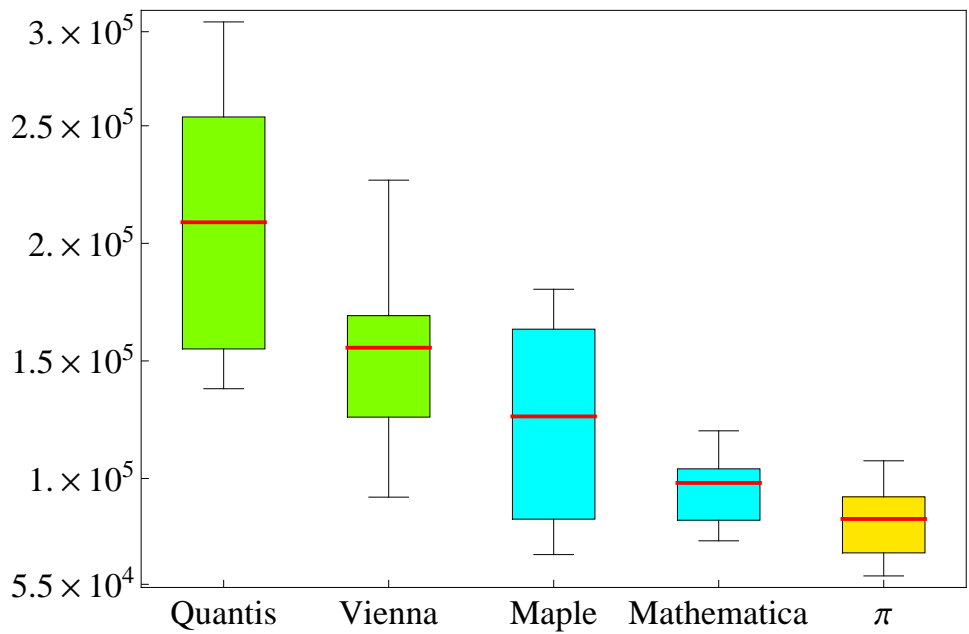}
   \caption{(Color online) Box-and-whisker plot for the results of the random walk tests.}
   \label{fig:example5}
\end{figure}

The fifth test
is thus based on viewing a  random
sequence as a one-dimensional {\em random walk;}
whereby the successive bits, associated  with an increase of one unit {\it per} bit of the $x$-coordinate,
are interpreted as follows: $1=$``move up,'' and  $0=$``move down'' on the $y$-axis.
In this way a measure is obtained for how far away one can reach from the starting point
(in either positive or negative) from the starting $y$-value of $0$ that one can reach using
successive bits of the sample sequence.
Figure~\ref{fig:example5} and Table~\ref{tab:5} summarize the results.

\begin{table}
\caption{Statistics for the results of the random walk tests.}\label{tab:5}
\begin{center}
\begin{tabular}
[c]{ c c c c c c c c }%
\hline\hline
Descriptive statistics & min & Q1 & median & Q3 & max & mean & sd\\\hline
Maple & 67640 & 88730 & 126400 & 162500 & 180500 & 125300 & 42995.59\\
Mathematica & 73500 & 84760 & 98110 & 103400 & 120300 & 96450 & 14685.34\\
Quantis  & 138200 & 161600 & 209000 & 250200 & 294200 & 211300 & 55960.23\\
Vienna & 92070 & 130200 & 155600 & 167600 & 226900 & 152900 & 36717.55\\
$\pi$ & 58570 & 70420 & 82800 & 91920 & 107500 & 82120 & 14833.75\\\hline\hline
\end{tabular}
\end{center}
\end{table}

\section{Statistical analysis of randomness tests results}

In what follows
the significance of results corresponding to each randomness test applied to all five sources have
been analyzed by means of some statistical comparison tests.
The Kolmogorov-Smirnov test  for two samples~\cite{Conover}
determines if two datasets differ significantly.
This test has the advantage of making no prior assumption about the distribution of data;
i.e., it is non-parametric and distribution free.

The Kolmogorov-Smirnov test returns a $p$-value,
and the decision ``the difference between the two datasets is statistically
significant" is accepted if the $p$-value {\em is less than $0.05$;}
or, stated pointedly, if the
probability of taking a wrong decision is less than $0.05$. Exact $p$-values are
only available for the two-sided two-sample tests with no ties.

In some cases we have tried to double-check the decision  ``no significant
differences between the datasets"  at the price of a supplementary, plausible
distribution assumption. Therefore, we have performed the Shapiro-Wilk test  for normality
\cite{Shapiro-Wilk} and, if normality is not rejected, we have assumed that the
datasets have normal (Gaussian) distributions.
In order to be able to compare the expected values (means) of the two samples,
the Welch $t$-test~\cite{Welch}, which is a version of Student's test, has been applied.
In order to emphasize the relevance of p-values less than 0.05 associated with Kolmogorov-Smirnov,
Shapiro-Wilk and Welch's $t$-tests, they are printed in boldface and discussed in the text.

\subsection{Book stack randomness test}

The results of the Kolmogorov-Smirnov test associated with the ``book-stack" tests are enumerated in Table~\ref{tab:8}.
Statistically significant differences are identified for
Quantis {\it versus} Mathematica and $\pi$.
\begin{table}
\caption{Kolmogorov-Smirnov test for the ``book-stack'' tests.}\label{tab:8}
 \begin{center}
 \begin{tabular}
[c]{ c c c c c }
\hline\hline
Kolmogorov-Smirnov test $p$-values & Mathematica & Quantis & Vienna & $\pi$ \\\hline
Maple  & 0.4175 & 0.1678 & 0.9945 & 0.4175\\
Mathematica &  & \bf{0.0021} & 0.1678 & 0.4175\\
Quantis &  &  & 0.1678 & \bf{0.0123}\\
Vienna &  &  &  & 0.4175\\\hline\hline
\end{tabular}
\end{center}
\end{table}

As more compression is a symptom of less randomness,  the corresponding ranking of
samples is as follows:
$\langle \text{Quantis} \rangle  = 89988.9 > \langle \text{Vienna} \rangle  = 53863.8  >  \langle \text{Maple} \rangle  =
53411.6 > \langle  \pi \rangle  = 41277.5  >  \langle \text{Mathematica} \rangle  = 27938.3$.
The Shapiro-Wilk tests results are presented in  Table~\ref{tab:9}.

\begin{table}
\caption{Shapiro-Wilk test for the ``book-stack'' tests.}\label{tab:9}
\begin{center}
\begin{tabular}
[c]{ c c c c c c }\hline\hline
Shapiro-Wilk test & Maple & Mathematica & Quantis & Vienna  & $\pi$\\\hline
$p$-value  & 0.7880 & 0.4819 & 0.7239 & 0.8146 &
0.5172\\\hline\hline
\end{tabular}
\end{center}
\end{table}

Since normality is not rejected for any string, we apply the Welch's $t$-test
for the comparison of means. The results are enumerated in Table~\ref{tab:10}.
Significant differences between the means are identified for the following sources:
(i) Quantis {\it versus} all other sources (Maple, Mathematica, Vienna, $\pi$); and
(ii)
Vienna {\it versus} Mathematica and Maple (as already mentioned).

\begin{table}
\caption{Welch's $t$-test for the ``book-stack'' tests.}\label{tab:10}
\begin{center}
\begin{tabular}
[c]{ c c c c c }\hline\hline
$p$-value & Mathematica & Quantis & Vienna & $\pi$\\\hline
Maple  & 0.0535 & \bf{0.0436} & 0.974 & 0.3412\\
Mathematica &  & \bf{0.0009} & \bf{0.0283} & 0.1551\\
Quantis  &  &  & \bf{0.0368} & \bf{0.0054}\\
Vienna &  &  &  & 0.2690\\\hline\hline
\end{tabular}
\end{center}
\end{table}

\subsection{Solovay-Strassen probabilistic primality test}

The Kolmogorov-Smirnov test results for this test are presented in Table~\ref{tab:11},
where no significant differences are detected.

The Shapiro-Wilk test results are presented in Table~\ref{tab:12}.
Since there is no clear
pattern of normality for the data, the application of Welch's $t$-test is not
appropriate.
\begin{table}
\caption{Kolmogorov-Smirnov test for the Solovay-Strassen tests.}\label{tab:11}
 \begin{center}
 \begin{tabular}
[c]{ c c c c c }
\hline\hline
Kolmogorov-Smirnov test $p$-values & Mathematica & Quantis & Vienna & $\pi$ \\\hline
Maple  & 0.7591 & 0.4005 & 0.7591 & 0.7591\\
Mathematica &  & 0.7591 & 0.7591 & 0.7591\\
Quantis &  &  & 0.4005 & 0.7591\\
Vienna &  &  &  & 0.9883\\\hline\hline
\end{tabular}
\end{center}
\end{table}

\begin{table}
\caption{Shapiro-Wilk test for the Solovay-Strassen tests.}\label{tab:12}
\begin{center}
\begin{tabular}
[c]{ c c c c c c }\hline\hline
Shapiro-Wilk test & Maple & Mathematica & Quantis  & Vienna  & $\pi$\\\hline
$p$-value & 0.0696 & \bf{0.0363} &
0.4378 & 0.6963 & 0.4315\\\hline\hline
\end{tabular}
\end{center}
\end{table}

\subsection{Borel test of normality}

The results of the Kolmogorov-Smirnov test are presented in Table~\ref{tab:5b}.
\begin{table}
\caption{Kolmogorov-Smirnov test for the Borel normality tests.}\label{tab:5b}
 \begin{center}
 \begin{tabular}
[c]{ c c c c c }
\hline\hline
Kolmogorov-Smirnov test $p$-values & Mathematica & Quantis & Vienna & $\pi$ \\\hline
Maple & 0.4175 &$\mathbf{< 10^{-4}}$
& \bf{0.0002}
& 0.1678\\
Mathematica &  & $\mathbf{< 10^{-4}}$
& \bf{0.0002} & 0.9945\\
Quantis &  &  & \bf{0.0002} & $\mathbf{< 10^{-4}}$
\\
Vienna &  &  &  & \bf{0.0002}\\\hline\hline
\end{tabular}
\end{center}
\end{table}
Statistically significant differences are identified for
(i) Quantis {\it versus}  Maple,  Maple, Mathematica and $\pi$;
(ii) Vienna {\it versus}  Maple, Mathematica and $\pi$; and
(iii) Quantis {\it versus}  Vienna.

Note that
\begin{enumerate}

\item Pseudo-random strings pass the Borel normality test for comparable, relatively small (with respect to quantum strings; cf. below),
numbers of counts:
if the angle brackets   $\langle x \rangle$ stand for the statistical mean of tests on $x$, then
 $\langle \text{Maple} \rangle  = 60210$, $\langle \text{Mathematica} \rangle  = 41870$,
$\langle  \pi \rangle  = 40220$).

\item Quantum strings pass the Borel normality test only for ``much larger numbers''
of counts
($\langle \text{Quantis} \rangle  = 207200$, $\langle \text{Vienna} \rangle  = 337100$).



\end{enumerate}
As a result, the Borel normality test detects and identifies
statistically significantly differences between all pairs of computable and incomputable  sources of ``randomness.''

\subsection{Test based on Shannon's information theory}

 The results of the Kolmogorov-Smirnov test are presented in  Table~\ref{tab:6}.
No significant differences are detected. The descriptive statistics
data for the results of this test  indicates almost identical
distributions corresponding to the five sources.

\begin{table}
\caption{Kolmogorov-Smirnov test for Shannon's information theory tests.}\label{tab:6}
 \begin{center}
 \begin{tabular}
[c]{ c c c c c }
\hline\hline
Kolmogorov-Smirnov test $p$-values & Mathematica & Quantis & Vienna & $\pi$ \\\hline
Maple & 0.7870 & 0.7870 & 0.7870 & 0.1678\\
Mathematica &  & 0.7870 & 0.4175 & 0.0525\\
Quantis &  &  & 0.4175 & 0.1678\\
Vienna &  &  &  & 0.4175\\\hline\hline
\end{tabular}
\end{center}
\end{table}

The results of the Shapiro-Wilk test associated with a test based on Shannon's information theory are presented in Table~\ref{tab:7}.
Since there is no clear
pattern of normality for the data, the application of Welch's $t$-test is not
appropriate.

\begin{table}
\caption{Shapiro-Wilk test for Shannon's information theory tests.}\label{tab:7}
\begin{center}
\begin{tabular}
[c]{ c c c c c c }\hline\hline
Shapiro-Wilk test & Maple & Mathematica & Quantis  & Vienna  & $\pi$\\\hline
$p$-value & 0.1962 & \bf{0.0189} &
\bf{0.0345} & 0.3790 & 0.8774\\\hline\hline
\end{tabular}
\end{center}
\end{table}

\subsection{Test based on random walks}

The Kolmogorov-Smirnov test results associated with test based on random walks are presented in Table~\ref{tab:13}.
Statistically significant differences are identified for:
(i) Quantis {\it versus} all other sources (Maple, Mathematica, Vienna and $\pi$);
(ii) Vienna {\it versus} Mathematica, Vienna (as already
mentioned) and  $\pi$;  and
(iii) Maple {\it versus} $\pi$.

Quantum strings move farther away from the starting point than the pseudo-random strings; i.e.,
$\langle \text{{\it Quantis}} \rangle > \langle \text{{\it Vienna}} \rangle > \langle \text{{\it Maple}} \rangle >  \langle \text{{\it Mathematica}} \rangle > \langle \pi \rangle  $.

\begin{table}
\caption{Kolmogorov-Smirnov test for the random walk tests.}\label{tab:13}
 \begin{center}
 \begin{tabular}
[c]{ c c c c c }
\hline\hline
Kolmogorov-Smirnov test $p$-values & Mathematica & Quantis & Vienna & $\pi$ \\\hline
Mathematica & 0.1678 & \bf{0.0123} & 0.4175 & 0.0525\\
Quantis &  & $\mathbf{< 10^{-4}}$ & \bf{0.0021} & 0.1678\\
Vienna &  &  & 0.0525 & $\mathbf{< 10^{-4}}$\\
$\pi$ &  &  &  & \bf{0.0002}\\\hline\hline
\end{tabular}
\end{center}
\end{table}

Note that quantum strings move farther away from the starting point than the pseudo-random strings; i.e.,
$\langle \text{{\it Quantis}} \rangle > \langle \text{{\it Vienna}} \rangle > \langle \text{{\it Maple}} \rangle >  \langle \text{{\it Mathematica}} \rangle > \langle \pi \rangle  $.
It was quite natural to double-check the conclusion  ``Quantis and Vienna do not
exhibit significant differences.'' Hence we run the Shapiro-Wilk test, which concludes
that normality is not rejected; cf. Table~\ref{tab:14}.

\begin{table}
\caption{Shapiro-Wilk test for the random walk tests.}\label{tab:14}
\begin{center}
\begin{tabular}
[c]{ c c c c c c }\hline\hline
Shapiro-Wilk test & Maple & Mathematica & Quantis & Vienna  & $\pi$\\\hline
$p$-value & 0.2006 & 0.9268 & 0.5464 & 0.8888 &
0.9577\\\hline\hline
\end{tabular}
\end{center}
\end{table}

Next, we apply the Welch's $t$-test for the comparison of means. The
results are given in Table~\ref{tab:15}.
Significant differences between the means are identified for the
following sources:
(i) Quantis {\it versus} all other sources (Maple, Quantis, Vienna, $\pi$);
(ii)
Vienna {\it versus} Mathematica), Quantis (as already mentioned) and $\pi$;
(iii) Maple {\it versus} $\pi$.

\begin{table}
\caption{Welch's $t$-tests for the random walk tests.}\label{tab:15}
\begin{center}
\begin{tabular}
[c]{ c c c c c }\hline\hline
$p$-value & Mathematica & Quantis & Vienna & $\pi$\\\hline
Maple & 0.06961 & \bf{0.0013} & 0.1409 & \bf{0.0119}\\
Mathematica &  & $\mathbf{< 10^{-4}}$ & \bf{0.0007} & \bf{0.0435}\\
Quantis &  &  & \bf{0.0143} & $\mathbf{< 10^{-4}}$\\
Vienna &  &  &  & \bf{0.0001}\\\hline\hline
\end{tabular}
\end{center}
\end{table}

\section{Summary}

Tests based on algorithmic information theory
analyze algorithmic randomness,  the strongest possible form of incomputability.
In this respect they differ from tests employed in the standard test batteries,
as the former depend on irreducible algorithmic information content,
which is constant for algorithmic pseudo-random generators.
Thus the set of randomness tests performed for our analysis could in principle be expected to be
``more sensitive'' with respect to differentiating
between quantum randomness and algorithmic types of ``quasi-randomness'' than statistical tests alone.

All tests have produced evidence --- with different degrees of statistical significance --- of differences between quantum and non-quantum sources.
In summary:

\begin{enumerate}
\item
For the test for Borel normality --- the strongest discriminator test --- statistically significant differences between the distributions of datasets are identified for
(i) {\it Quantis} versus {\it Maple}, {\it Mathematica} and $\pi$; (ii) {\it Vienna} versus {\it Maple}, {\it Mathematica} and $\pi$; and (iii) {\it Quantis} versus {\it Vienna}.

Not only that the average number of counts is larger for quantum sources, but the
increase is quite significant: {\it Quantis}  is $3.5-5$ times larger than the corresponding average number
of counts for software-generated sources, and {\it Vienna} is $5-8$ times larger than
those values.
\item
For the test based on random walks, statistically significant differences between the distributions of datasets are identified for:
(i) {\it Quantis} versus all other sources ({\it Maple}, {\it Mathematica}, {\it Vienna} and $\pi$);
(ii) {\it Vienna} versus {\it Mathematica}, {\it Vienna} and $\pi$. Quantum strings move farther away from the starting point than the pseudo-random strings; i.e.,
$\langle \text{{\it Quantis}} \rangle > \langle \text{{\it Vienna}} \rangle > \langle \text{{\it Maple}} \rangle >  \langle \text{{\it Mathematica}} \rangle > \langle \pi \rangle  $.

\item
For the ``book-stack'' test, significant differences between the means are identified for the following sources: (i) {\it Quantis} versus all other sources ({\it Maple}, {\it Mathematica}, {\it Vienna}, $\pi$);
and (ii) {\it Vienna} versus {\it Mathematica} and {\it Maple}.

\item
For the test based on Shannon's information theory, as well as for the Solovay-Strassen test, {\em no significant differences} among the five chosen sources are detected. In the first case the reason may come from the fact  that averages are the same for all samples. In the second case  the reason  may be due to the fact that the test is based solely on the behavior of algorithmic random strings and not on a specific property of randomness.

\end{enumerate}

We close with a cautious remark about the impossibility to formally or experimentally ``prove absolute randomness.''
Any claim of randomness can only be secured  {\em relative} to, and
{\em with respect} to, a more or less large class of laws or behaviors, as
it is impossible to inspect the hypothesis against an infinity of --- and even less so all --- conceivable laws.
To rephrase a statement about computability~\cite[p. 11]{davis-58}, {\em ``how can we ever exclude the possibility of our
presented, some day (perhaps by some extraterrestrial visitors), with a (perhaps
extremely complex) device  that ``computes'' and ``predicts''
a certain type of hitherto ``random'' physical behavior?''}

\section*{Acknowledgements}

We are grateful to  Thomas Jennewein and Anton Zeilinger for providing us with the quantum random bits produced at the University of  Vienna by  Vienna IQOQI group, for the description of their method, critical comments and  interest in this research.

We thank: a) Alastair Abbott, Hector Zenil and Boris Ryabko for interesting comments, b)
Ulrich Speidel for his tests for which some partial results have been reported in our extended paper~\cite{CDMTCS372},
c) Stefan Wegenkittl for critical comments of various drafts of this paper and his suggestions  to exclude  some tests,
d) the anonymous Referees for constructive suggestions.

Calude gratefully  acknowledges the support of  the Hood Foundation  and the  Technical University of  Vienna. Svozil  gratefully  acknowledges support of the CDMTCS at the University of Auckland, as well as of the Ausseninstitut of the Vienna University of Technology.


%

\end{document}